\documentclass[aps,nofootinbib,preprint]{revtex4}

\usepackage{graphicx}

\usepackage{amsmath}
\usepackage{amsfonts}
\usepackage{amssymb}
\usepackage{hyperref}
\usepackage{color}

\newcommand{\be}{\begin{equation}}
\newcommand{\ee}{\end{equation}}

\def\beq{\begin{equation}}
\def\eeq{\end{equation}}
\def\bea{\begin{eqnarray}}
\def\eea{\end{eqnarray}}

\def\slashchar#1{\setbox0=\hbox{$#1$}           
   \dimen0=\wd0                                 
   \setbox1=\hbox{/} \dimen1=\wd1               
   \ifdim\dimen0>\dimen1                        
      \rlap{\hbox to \dimen0{\hfil/\hfil}}      
      #1                                        
   \else                                        
      \rlap{\hbox to \dimen1{\hfil$#1$\hfil}}   
      /                                         
   \fi}

\begin{document}

\vspace*{1cm}
\title{When CoGeNT met PAMELA}

\author{\vspace{0.5cm}Wai-Yee Keung$\,^a$,  Ian Low$\,^{b,c}$,  Gabe Shaughnessy$\,^{b,c}$}
\affiliation{\vspace{0.5cm}
\mbox{$^a$ Department of Physics, University of Illinois, Chicago, IL 60607}\\
\mbox{$^{b}$High Energy Physics Division, Argonne National Laboratory,
Argonne, IL 60439}\\ 
\mbox{$^{c}$Department of Physics and Astronomy, Northwestern University,
Evanston, IL 60208}\\
\vspace{0.8cm}
}

\begin{abstract}
\vspace*{0.2cm} 
If the excess events from the CoGeNT experiment arise from elastic scatterings of a light dark matter off the nuclei, crossing symmetry implies non-vanishing annihilation cross-sections of the light dark matter into hadronic final states inside the galactic halo, which we confront with the anti-proton spectrum measured by the PAMELA collaboration. We consider two types of effective interactions between the dark matter and the quarks: 1) contact interactions from integrating out heavy particles and 2) long-range interactions due to the electromagnetic properties of the dark matter. The lack of excess in the anti-proton spectrum results in tensions for a scalar and, to a less extent, a vector dark matter interacting with the quarks through the Higgs portal. 
\end{abstract}

\maketitle


\section{Introduction}

In recent years we have witnessed a plethora of experimental anomalies in both direct and indirect detection experiments in search of the dark matter (DM). For example, the direct detection experiment DAMA \cite{Bernabei:2008yi} observed signals of annual modulations consistent with that expected from dark matter scattering off the nuclei, while the CDMS collaboration announced two candidate events for detection of dark matter~\cite{CDMS}. More recently, the CoGeNT experiment \cite{Aalseth:2010vx} released results containing signal events suggestive of a light dark matter in the range of 7 -- 12 GeV. On the other hand, several collaborations in indirect detections reported excessive fluxes of electrons/positrons and photons, such as the PAMELA \cite{Adriani:2008zr}, the ATIC \cite{ATIC:2008zzr}, the Fermi LAT \cite{Abdo:2009zk}, the HESS \cite{Collaboration:2008aaa}, and the WMAP \cite{Finkbeiner:2004us}, which may not be explained away easily by conventional astrophysical sources. 

In order to understand whether these experimental excesses are indeed the prelude to a discovery of the dark matter, it is crucial to cross-check dark matter interpretations of these empirical anomalies against one another. One possibility is to construct a unified theory explaining all these signal events \cite{ArkaniHamed:2008qn}. Another  is to confront the events with constraints from measurements which do not see any excess, one example of which is the anti-proton spectrum measured also by the PAMELA \cite{Adriani:2008zq}. While this has been generally been studied for Dirac DM with a contact interaction~\cite{Cao:2009uv} and specifically for 10 GeV DM~\cite{Lavalle:2010yw}, In this work we propose to study the consistency of the CoGeNT signals with the anti-proton spectrum of PAMELA within the context of generic effective interactions between DM and quarks.  

If the signal events in various direct detection experiments are due to elastic scatterings of the dark matter off the nuclei, which arise from the dark matter interaction with the quarks, then crossing symmetry implies the same interactions must also give rise to annihilations of dark matter into quarks. This argument has been used to  constrain the boost factor \cite{Hisano:2004ds} necessary to explain the electron/positron excess in the PAMELA, using the updated limits from the CDMS \cite{Cao:2009uv}. Another example studied the implication of various direct detection bounds on the anti-deuteron flux from hadronic annihilations of the dark matter \cite{Cui:2010ud}.

Theoretically, the light ( 7 -- 12 GeV) dark matter interpretation of the CoGeNT offers a unique opportunity for a model-independent study. Since such a mass scale is far below the electroweak scale, it seems reasonable to assume that any degrees of freedom mediating the dark matter interactions with the standard model particles  have masses much heavier than that of the dark matter.\footnote{There are exceptions to this assumption. See, for example, Ref.~\cite{Draper:2010ew}.}  Hence at  energy scales relevant for the elastic scattering of dark matter with the nucleons and the dark matter annihilations into quarks, these ``mediator particles'' can be integrated out and the interactions  could be approximated by higher dimensional operators. The resulting effective interactions can be either contact interactions, when the mediator particle interacts with the quarks, or long range in nature, when the mediator particle interacts with the photon.

By combining the PAMELA anti-proton spectrum and the putative CoGeNT signal, we find that contact interactions of a scalar or vector dark matter, arising from a heavy mediator particle with Yukawa-like interactions with quarks, is the most severely constrained.   These constraints apply to models with a light scalar dark matter which interacts with the standard model through the Higgs portal \cite{HiggsPortal}. Cases of fermionic dark matter with an electromagnetic charge form factor coupling or magnetic moment coupling also exhibit tension with these data, but to a lesser degree. Previous studies on properties of the dark matter using higher dimensional contact interactions can be found in Refs.~\cite{Beltran:2008xg, Shepherd:2009sa}.

This work is organized as follows: we first classify the effective interactions relevant for this study in Section \ref{sect:II}. Expressions for the scattering and annihilation cross-sections are given in Section \ref{sect:III}. Then we present the results in Section \ref{sect:results}, followed by the conclusion.

\section{Classification of effective interactions}
\label{sect:II}

There are two possibilities for the dark matter effective interactions upon integrating out the heavy degrees of freedom: the mediator particle could induce either direct contact interactions between the dark matter and the quarks or electromagnetic interactions of the dark matter, which would then interact with the quarks through the photon.

We are interested in operators which could serve as the leading order contribution to both the spin-independent (SI) scattering cross-sections in direct detections and the annihilation cross-sections in indirect detections. Since both the scattering and the annihilation processes are non-relativistic, we will focus on operators that are not suppressed by the velocity of either the dark matter or the nucleon. For indirect detections this implies we only consider $s$-wave annihilations. The list of operators satisfying the above two conditions turn out to be quite short, as we explain below.

\subsection{Contact Interactions}

The quark bilinears contributing to the SI elastic scattering without velocity suppressions have only two forms  \cite{Agrawal:2010fh,Cui:2010ud, Fan:2010gt}:
$ \bar{q}\gamma^\mu q$  and $ \bar{q}q$,
where the second bilinear violates the chiral symmetry and will dominate only when the first contribution is absent. Much of what appears below can be found in Refs.~\cite{Agrawal:2010fh,Cui:2010ud, Fan:2010gt, Beltran:2008xg}.

\begin{itemize}

\item  For a scalar dark matter $\phi$ the only operator utilizes the $\bar{q}q$ bilinear:
\be
\label{eq:os}
{\cal O}_{s} = c_s^q\, \frac{2 m_\phi}{\Lambda_s^2} |\phi|^2 \,
\bar{q}q \ .  \ee 
The factor $2m_\phi$ mimics a similar flux normalization in the
non-relativistic limit as in the
case of fermionic dark matter.  In this way, a universal expression for
SI cross-sections can be given in the following section.
However, with this particular normalization, the parameter $\Lambda_s$ may not directly
reflect the underlying short distance scale. Their relation  requires 
further knowledge of the short distance physics involved.
If the dark matter is a real scalar, simply replace $|\phi|^2 \to
\phi^2$. The operator that invokes $\bar{q}\gamma^\mu q$ bilinear, $\phi^*
\tensor{\partial}_\mu \phi\, \bar{q}\gamma^\mu q$, only leads to
$p$-wave annihilations\footnote{The operator,
$\partial_\mu(\phi^*\phi)\, \bar{q}\gamma^\mu q$, vanishes by the equation of motion upon integration by parts.}.

\item For a fermionic dark matter $\chi$, there is only one  operator
contributing to  both SI scattering and $s$-wave annihilation without
velocity suppression:
\be
{\cal O}_{f} = 
\frac{c_{f}^q}{\Lambda_{f}^2} \bar{\chi}\gamma^\mu\chi\,\bar{q}\gamma_\mu q
 \  ,
\ee
hereafter referred to as the Dirac contact model.  Note that  this operator only applies to a Dirac fermion, otherwise it is exactly zero for a Majorana fermion.
Two operators contributing to the SI scattering, $\bar{\chi}\gamma^5\chi\,
\bar{q} q$ and $\bar{\chi}\gamma^\mu\gamma^5 \chi\, \bar{q}\gamma_\mu q$, are velocity-suppressed, while the operator, $\bar{\chi}\chi\,
\bar{q} q$, only annihilates through $p$-wave.

\item For a vector dark matter $B'_\mu$, again there is only one operators:
\be
\label{eq:ov}
{\cal O}_{v}= c_{v}^q\,\frac{2m_{B'}}{\Lambda_{v}^2} B^{\prime \, \dagger}_\mu B^{\prime\,\mu} \, \bar{q} q  \ .
\ee
The discussion for a vector dark matter is identical to that of a scalar dark matter, as ${\cal O}_v$ and ${\cal O}_s$ are quite similar. In addition, the other two operators contributing to SI cross-sections,  $\partial_\mu( B'^\dagger_\nu  B^{\prime\, \nu}) \, \bar{q}\gamma^\mu q$ and $B'^\dagger_\nu \tensor{\partial}_\mu B^{\prime\, \nu} \, \bar{q}\gamma^\mu q$, either vanishes by the equation of motion (the former) or only annihilates through $p$-wave (the latter).

\end{itemize}

\subsection{Electromagnetic Interactions}

Here we assume the dark matter is a Dirac fermion and consider the
possibility that the dark matter may possess electromagnetic multipole
moments or a charge form factor  \cite{Pospelov:2000bq,Barger:2010gv,Fitzpatrick:2010br,Banks:2010eh},
giving rise to interactions with the quarks through the photon. At
dimension-five level, the electric  dipole moment (EDM) and the
magnetic dipole moment (MDM) are
\be
{\cal O}_{E} = \frac12 \frac{e}{\Lambda_E} 
           \bar{\chi}\sigma^{\mu\nu}\gamma^5 \chi\, {F}^{\mu\nu} \ , \qquad
{\cal O}_{M} = \frac12 \frac{e}{\Lambda_M} 
          \bar{\chi}\sigma^{\mu\nu}\chi\, F^{\mu\nu} \ .
\ee
It turns out the operator ${\cal O}_E$ annihilates through the $p$-wave, which can be seen by considering the charge conjugation $(C)$ and parity $(P)$ properties as follows \cite{Cui:2010ud}:
\begin{itemize}
\item
The $C$ and $P$ properties of an initial state with two fermions are given by $C=(-1)^{L+S}$ and $P=(-1)^{L+1}$, while those of the fermion bilinear $\bar{\chi}\sigma^{\mu\nu}\gamma^5 \chi$ are $C=-1$ and $P=-(-1)^{\mu}(-1)^{\nu}$, where $(-1)^\mu=+1$ for $\mu=0$ and $-1$ for $\mu=1,2,3$.  The $s$-wave $(L=0)$ annihilation requires $P=-1$, which together with $C=-1$ of the bilinear determines the initial state to be in the $S=1$ and $J=L+S=1$ state.

\item 
The condition $P=-1$ for the $s$-wave and the identity  $\sigma^{00}=0$ select the  $\sigma^{ij}\gamma^5$ components of the fermion bilinear, which in turn single out the $F^{ij}$ components of the field strength tensor. However, in momentum space $F^{ij}$ are proportional to the 3-momentum of the virtual photon, which is zero in the centre-of-mass frame.  Therefore the annihilation must occur through the $p$-wave state.
\end{itemize}

At dimension-six level there are the charge form factor (CFF)
operator and the anapole operator (A):
\be
{\cal O}_{C} = \frac{e}{\Lambda_C^2}\, \bar{\chi}\gamma_\mu\chi\ 
                          \partial_\nu F^{\mu\nu} 
\ ,\qquad
{\cal O}_{A} = \frac{e}{\Lambda_A^2}\, \bar{\chi}\gamma_\mu\gamma^5\chi\ 
                           \partial_\nu F^{\mu\nu} \ .  
\ee
Note that other forms constructed from the field strength dual are zero
because of the absence of magnetic monopole current
$\partial_\nu \widetilde{F}^{\mu\nu}=0$.
The operator ${\cal O}_{C}$  gives rise to a
Feynman vertex $\bar\chi\gamma_\mu\chi$ with  the form factor
$G_\chi(\mathbf{q}^2)$,  parametrized in Ref.~\cite{Barger:2010gv} as
\be
G_\chi(\mathbf{q}^2) = \frac{\mathbf{q}^2}{\Lambda_C^2} \ ,
\ee
which vanishes as $\mathbf{q}^2 \to 0$ as required by that  the neutrality of the dark matter.
Using the  equation of motion, 
$\partial_\nu F^{\mu\nu}=e \sum_q e_q \bar q \gamma^\mu q $, we can
turn ${\cal O}_{C}$ and ${\cal O}_{A}$  into contact interactions,
\be
 {\cal O}_C \longrightarrow  
    \frac{e^2}{\Lambda_C^2}\, \bar{\chi}\gamma_\mu\chi
       \sum_q e_q\bar q\gamma^\mu q \ ,\qquad
 {\cal O}_A \longrightarrow 
    \frac{e^2}{\Lambda_A^2}\, \bar{\chi}\gamma_\mu\gamma^5\chi
        \sum_q e_q\bar q\gamma^\mu q \ .
\ee
Then  ${\cal O}_{A}$ gives rise to velocity-suppressed SI cross-sections and need not be considered further.
 
To summarize, in this category we only consider two operators: ${\cal O}_M$ and ${\cal O}_{C}$.

\section{Cross-sections}
\label{sect:III}

Here we collect the formulas for the cross-sections relevant for the direct detection experiments such as CoGeNT and the indirect detection experiments such as the PAMELA, listed by the operators discussed in Section \ref{sect:II}.

\subsection{Direct Detections}
For contact interactions, the SI cross-section of elastic dark matter-nucleus
scattering can be written as 
\be 
\label{eq:direct}
\sigma_0 = \frac{\delta_m}{\pi} m_{r}^2 [ Z f_p + (A-Z) f_n]^2\ , 
\ee
where $\delta_m=4$ if the dark matter is a self-conjugated field, such as a real scalar, a
Majorana fermion, or a real vector field, while $\delta_m=1$ in  other cases.  The factor of 4 originates from two equivalent Wick contractions with the initial and final states when the dark matter is self-conjugate.  
Also $m_r$ is the reduced mass of the
dark matter-nucleus system, $A$ and $Z$ are the atomic mass and atomic
number of the target nuclei, respectively, and $f_{n,p}$ are the
effective couplings of the dark matter to the proton and the
neutron. 

With our choice of normalization for the couplings in
Eqs.~(\ref{eq:os}--\ref{eq:ov}), we can write
\bea
{\cal O}_{s,v}:&\quad&  f_{n,p} = \sum_{q=u,d,s} \frac{c_{s,v}^q}{\Lambda_{s,v}^2}
\frac{m_{n,p}}{m_q} f^{(n,p)}_{Tq} + \frac2{27}\left(1-\sum_{q=u,d,s}
f^{(n,p)}_{Tq}\right) \sum_{q=c,b,t}
\frac{c_{s,v}^q}{\Lambda_{s,v}^2}\frac{m_{n,p}}{m_q} \  \label{eq:fs} \\ 
{\cal O}_f: &\quad& f_{n} = \frac{1}{\Lambda_f^2}(c_f^u+2 c_f^d) \ , \quad f_p =
\frac{1}{\Lambda_f^2} (2 c_f^u+ c_f^d)  \ .
\label {eq:fv}
\eea
In Eq.~(\ref{eq:fs}) the first  grouping is the contributions of the light quarks $(u, d, s)$, while 
the second grouping is the gluonic contributions 
induced by the heavy quarks $(c, b, t)$, and  
$f^p_{Tq}m_p\equiv \langle p| m_q\bar q q|p\rangle$
effectively represents  the light quark  fraction of the proton mass.
Their numerical values for the proton and the neutron have been
estimated to be \cite{Ellis:2000ds} 
\bea
f^{p}_{Tu}&=&0.020\pm0.004\ , \quad f^{p}_{Td}=0.026\pm0.005 \ , \quad f^{p}_{Ts}=0.118\pm0.062 \ , \nonumber \\
f^{n}_{Tu}&=& 0.014\pm0.003\ ,\quad f^{n}_{Td}=0.036\pm0.008\ , \quad f^{n}_{Ts}=0.118\pm0.062 \ .
\eea
The value of $1-\sum_{u,d,s}f^{(p,n)}_{Tq}$  is approximately
$0.84$ and $0.83$ for protons and neutrons, respectively.
For the case of ${\cal O}_f$, 
$f_p$ and $f_p$ directly count the up quark or the down quark
components in the proton and the neutron, respectively.  Note that the large uncertainty in the strange-quark contribution imparts an ${\cal O}(30\%)$ uncertainty in the scattering cross section \cite{Barger:2008qd}, but does not affect our conclusions greatly.  Furthermore, we will use the following fermion masses:
\bea
&&m_u = 0.002\, {\rm GeV} \ , \quad m_d = 0.005\, {\rm GeV} \ , \quad m_s = 0.095\, {\rm GeV}\ , \nonumber \\
&&m_c=1.25\phantom{0}\,\, {\rm GeV} \ , \quad m_b = 4.2\phantom{01} \,\, {\rm GeV} \ , \quad m_t=172.3\, {\rm GeV} \  , \nonumber \\
&& m_n=0.939\, {\rm GeV} \ , \quad m_p=0.938\, {\rm GeV} \ . \nonumber
\eea

It is worth commenting that, since the quark bilinear involved in ${\cal O}_{s,v}$ breaks chiral symmetry, one could consider two possibilities for the couplings $c_{s,v}$: the non-universal coupling, for which $c_{s,v}^q=\tilde{c}_{s,v} \sqrt{2} m_q/v$ is proportional to the Yukawa coupling, and the universal coupling. The effective dark matter-nucleon couplings are very different:
\bea
{\rm Universal}:&& f_n = 14.56 \,\frac{c_{s,v}}{\Lambda_{s,v}^2} \ , \qquad \qquad \, \, \, \, f_p= 15.48\, \frac{c_{s,v}}{\Lambda_{s,v}^2} \ , \\
{\rm Non-universal}:&& f_n = 1.91\times 10^{-3}\, \frac{\tilde{c}_{s,v}}{\Lambda_{s,v}^2} \ , \qquad f_p= 1.89\times 10^{-3} \,\frac{\tilde{c}_{s,v}}{\Lambda_{s,v}^2} \ .
\eea

The actual cross-section including the nuclear form factor effect can be written in terms of $\sigma_0$:
\be 
\label{eq:sigma0}
\frac{d\sigma}{d\mathbf{q}^2} =
\frac{\sigma_0}{4m_r^2 v_r^2} F^2(\mathbf{q}^2) \ , 
\ee 
where $0< \mathbf{q}^2 < (2 m_r v_r)^2$. Here $v_r$ is the relative velocity between the dark matter and the target nuclei. The nuclear recoil energy $E_r$ is related to $\mathbf{q}^2=2 m_N E_r$. Notice that $\sigma_0$ is the number that is usually quoted in direct detection experiments.

For the MDM contribution, we use the result and the notation from Ref.~\cite{Barger:2010gv}:
\be
\label{eq:mdmsigma}
 {d\sigma\over d \mathbf{q}^2}
={e^4 \over 4\pi \Lambda_M^2\mathbf{q}^2 } 
\left[ Z^2 -{ Z^2\mathbf{q}^2\over 4 m_N^2  v_r^2}
     \left( 1 -{2m_N\over  m_{\rm DM}}\right) 
      + {I+1\over 3I}\left( {\mu_{Z,A}\over {e\over 2m_p}} \right)^2
{\mathbf{q}^2 \over 2 m_p^2 v_r^2} 
\right] F^2(\mathbf{q}^2) 
\ .\ee
Because of the long range interaction, $\sigma_0$ as defined in Eq.~(\ref{eq:sigma0}) is divergent and one must compare the nuclear recoil energy with experiments directly. In Eq.~(\ref{eq:mdmsigma}) the leading nuclear charge effect of the $Z^2$ term, though not associated with $1/v_r^2$, receives large enhancement from $1/\mathbf{q}^2$ from the virtual photon propagator. For simplicity, we have assumed the same nucleus form factor $F$ for the charge $Z^2$ and for the moment $\mu_{Z,A}$~\cite{Barger:2010gv}
\be
F(q) = {4 \pi^2 \rho a_0 c \over q} \left[{\pi a_0\over c} {\sin(q c) \cosh(\pi a_0 q)\over \sinh^2(\pi a_0 q)} - {\cos(q c)\over \sinh(\pi a_0 q)}\right],
\ee
where $a_0=2.9\text{ GeV}^{-1}$ and $c=(5.98 A^{1/3}-2.43) \text{ GeV}^{-1}$ are the characteristic scales, and $\rho_0 = {3\over 4 \pi c^3} {1\over 1+ a_0^2 \pi^2/c^2}$ is the normalized density of the Fermi distribution $\rho({\bf r})=\rho_0 / (1+e^{(r-c)/a_0})$.

\subsection{Indirect Detection}

\begin{table}[t]
\begin{tabular}{|c|c|c|c|c|} \hline
 \makebox[1cm]{ }& \makebox[1.5cm]{${\cal O}_s$} & \makebox[1.5cm]{${\cal O}_f$} & \makebox[1.5cm]{${\cal O}_v$} & \makebox[1.5cm]{${\cal O}_M$}
\\ \hline \hline
$a_i$ & 0 & $3/2$ & 0 & $3/8$  \\ \hline
$b_i$ & 1 & $-1/2$ & $1/3$ & $-1/8$ \\ \hline
\end{tabular}
\caption{\label{table1}\em Annihilation cross-sections for various operators, using the expression in Eq.~(\ref{eq:anni}).}
\end{table}

The thermally averaged annihilation cross-sections for the effective operators
into hadronic states in the non-relativistic $v \to 0$ limit are given by
\be
\label{eq:anni}
(\sigma v)_{\rm an}=\frac{N_c \delta_m}{\pi}
\sum_q \left(\frac{c_i^q}{\Lambda_i^2}\right)^2 
m_{\rm DM}^2 \sqrt{1-m_q^2/m_{\rm DM}^2}
 \left[ a_i + b_i \left(1-\frac{m_q^2}{m_{\rm DM}^2}\right)\right] \ ,
\ee
where the values of $a_i$ and $b_i$ are listed in Table
\ref{table1}. The color factor $N_c$ is 3 for the quark. Notice that
the annihilation cross-section for the vector dark matter is one third
of that for the scalar dark matter. However, the relevant annihilation
cross-section for the thermal relic abundance and the annihilation
rate in the galaxy is not $(\sigma v)_{\rm an}$; 
instead it is 
\be
\langle \sigma v\rangle_{\rm eff} = {\delta_b}\, (\sigma v)_{\rm an} \ ,
\ee
where $\delta_b=1/2$ if the dark matter is a complex field, such as the complex scalar or the Dirac fermion, and 1 otherwise \cite{Srednicki:1988ce}. The factor of 1/2 accounts for the fact that only two out of the four possible annihilation channels are non-zero ($\chi\bar{\chi}$ and $\bar{\chi}\chi$ but not ${\chi}\chi$ and $\bar{\chi}\bar\chi$.)

It is worth commenting that $\delta_m$ is common in Eq.~(\ref{eq:direct}) and in Eq.~(\ref{eq:anni}). Therefore the annihilation rate predicted from crossing symmetry, using the measured elastic scattering cross-section, for a complex dark matter is a factor of $\delta_b=1/2$ smaller than that from a self-conjugate dark matter. Hence we only focus on the case of real dark matter when presenting our results in Section \ref{sect:results}.

For the annihilation into $q\bar q$ via the MDM operator, simply make the following replacement in Eq.~(\ref{eq:anni}):
\be
\left(\frac{c_i^q}{\Lambda_i^2}\right)^2 
m_{\rm DM}^2  \to e_q^2 \left({e^2 \over\Lambda_M}\right)^2 \ .
\ee
Also for ${\cal O}_M$  the annihilation formula is a good approximation if $m_{\rm DM}\ll m_Z$, which is the case of interest for CoGeNT results.\footnote{The effect of the $Z$-boson could be included in a straightforward way \cite{Heo:2009vt}. }

\section{Results}
\label{sect:results}

Due to crossing symmetry, the possible  dark matter scattering signal may be converted into a possible dark matter annihilation signal and vice versa~\cite{Cao:2009uv}, if we assume both signals originate from the same interaction.  This allows one to map, using a well defined model or effective interaction, the scattering cross-section measured by direct detection experiments such as CoGeNT or XENON into a region in the plane of the dark matter annihilation cross-section.  Since dark matter-nucleon scattering occurs via quarks, crossing symmetry allows one to compute the corresponding annihilation cross-section into hadronic final states.  We follow this procedure to obtain the prediction of $(\sigma v)_{\rm an}$ for all models we consider. Special care must be taken for the MDM model, where the non-standard ${\bf q}^2$ dependence requires us to explicitly fit the CoGeNT recoil spectrum \cite{Barger:2010gv}.

The annihilation into quarks can imprint features onto the $\bar p/p$
spectra as measured by the satellite PAMELA~\cite{Adriani:2008zq}.
The $\bar p/p$ spectrum measured by PAMELA in the 1-100 GeV range with
500 days of running shows no significant deviation from the background expected from astrophysical cosmic ray sources.  This agreement with the expected background can serve to limit the total annihilation rate of dark matter to hadronic final states \cite{Cirelli:2008pk}.  

\begin{figure}[t]
\includegraphics[scale=0.49]{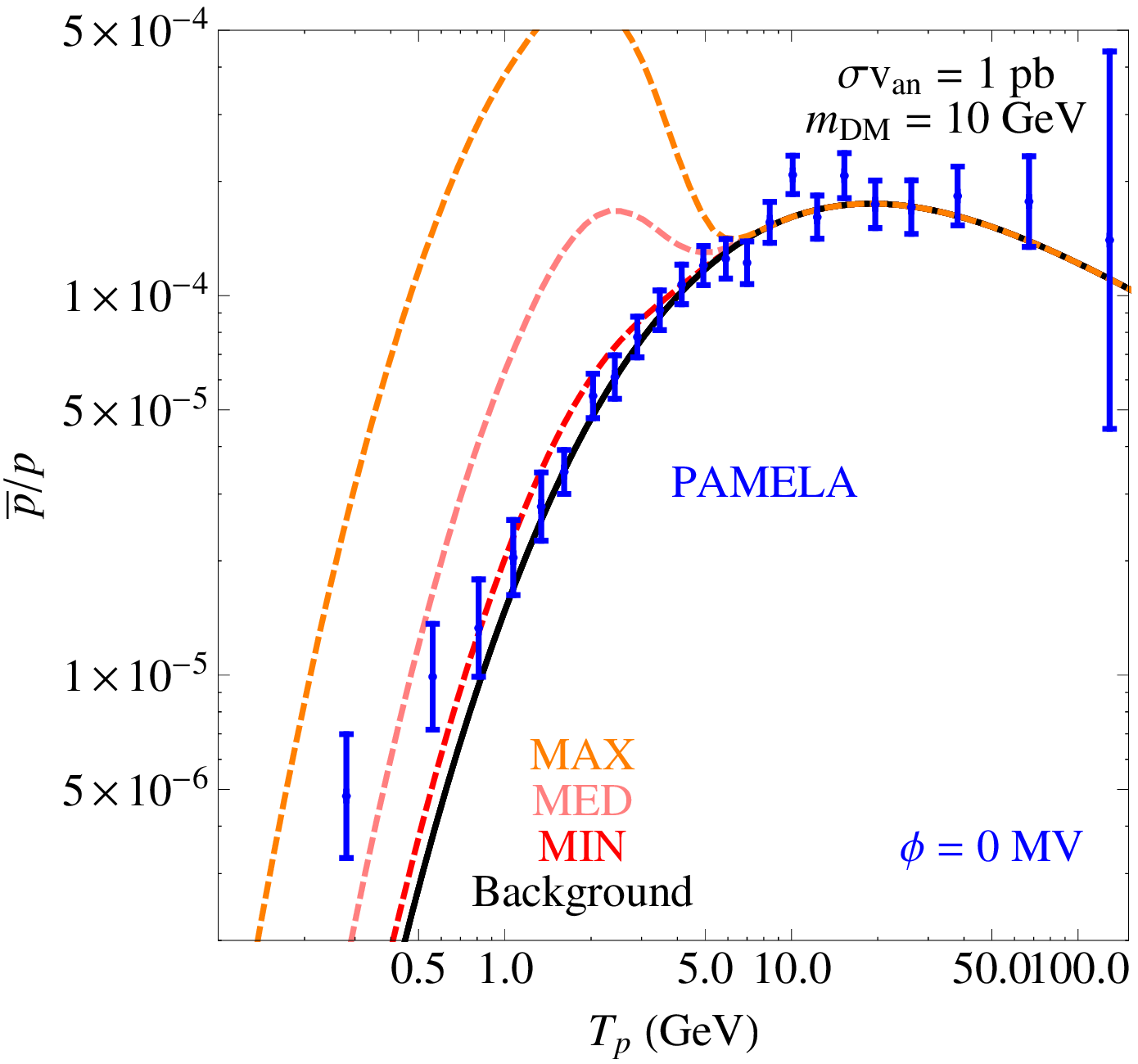} 
\includegraphics[scale=0.49]{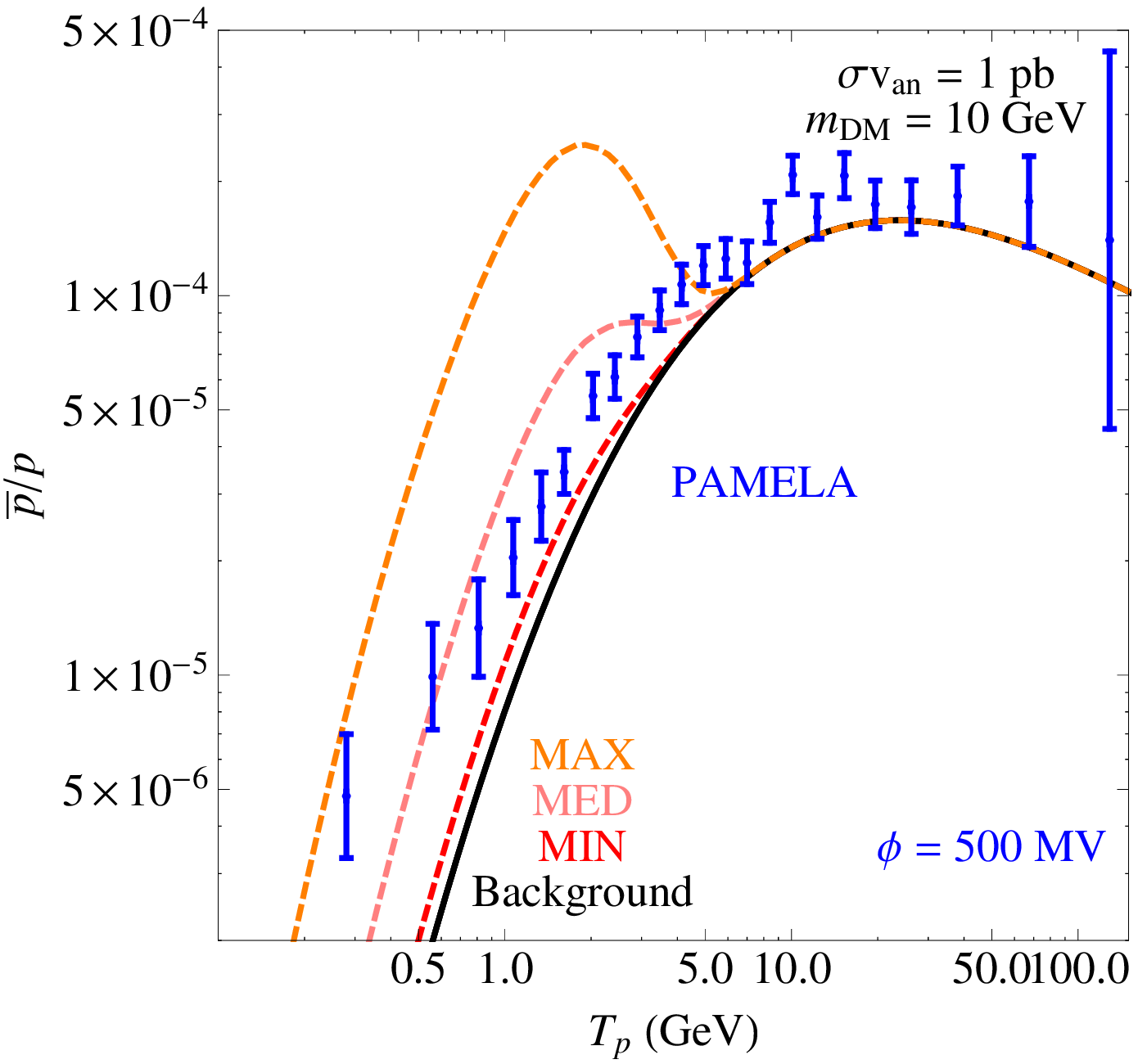} 
\caption{\em PAMELA $\bar p/p$ data overlain with the resulting $\bar p/p$ spectra from 10 GeV dark matter annihilating to quarks with cross section $(\sigma v)_{\rm an} = 1$ pb and solar modulation potential of $\phi = 0$ MV and $\phi = 500$ MV, respectively.  The large range of uncertainty in the $\bar p$ flux from the propagation model is evident. \em \label{fig:fig1} }
\end{figure}

It is important to emphasize that there are large uncertainties in the total flux of cosmogenic anti-protons due to the variations in the $\bar p$ propagation model.  We adopt three propagation models that give the maximal, median and minimal $\bar p$ flux compatible with an analysis of the Boron to Carbon ratio found in cosmic rays, and are denoted MAX, MED and MIN, respectively.  Therefore, these three models can be viewed as a measure of the uncertainty of the $\bar p$ flux due to the propagation model.  The details of the anti-proton propagation model are specified in Ref.~\cite{Donato:2003xg}, with the MAX, MED, and MIN scenarios defined in its Table ~\label{table1}. Further details related to these propagation models appear in Refs.~\cite{Delahaye:2007fr,Cirelli:2008id}.  To compute the $\bar p / p$ and apply the PAMELA constraint, we adopt the analysis given in Ref.~\cite{Barger:2008su} and include the effect of charge dependent solar modulation described in Ref.~\cite{Cirelli:2008id}.  In Fig.~\ref{fig:fig1}, we show the PAMELA $\bar p/p$ ratio data along with the spectra of injected $\bar p$ due to annihilations of a 10 GeV dark matter with a total cross section of $(\sigma v)_{\rm an} = 1$ pb, assuming equal coupling to all quarks and the Navarro-Frenk-White DM halo profile~\cite{NFW}.  We illustrate the effect solar modulation has by comparing the no modulation case, $\phi = 0$ MV, and modulation representative of the solar minimum, $\phi = 500$ MV, when the PAMELA data was acquired.  It is evident the uncertainty from the propagation model manifests a large range of $\bar p$ flux, spanning more than an order of magnitude.  

\begin{figure}[t]
\includegraphics[scale=0.49]{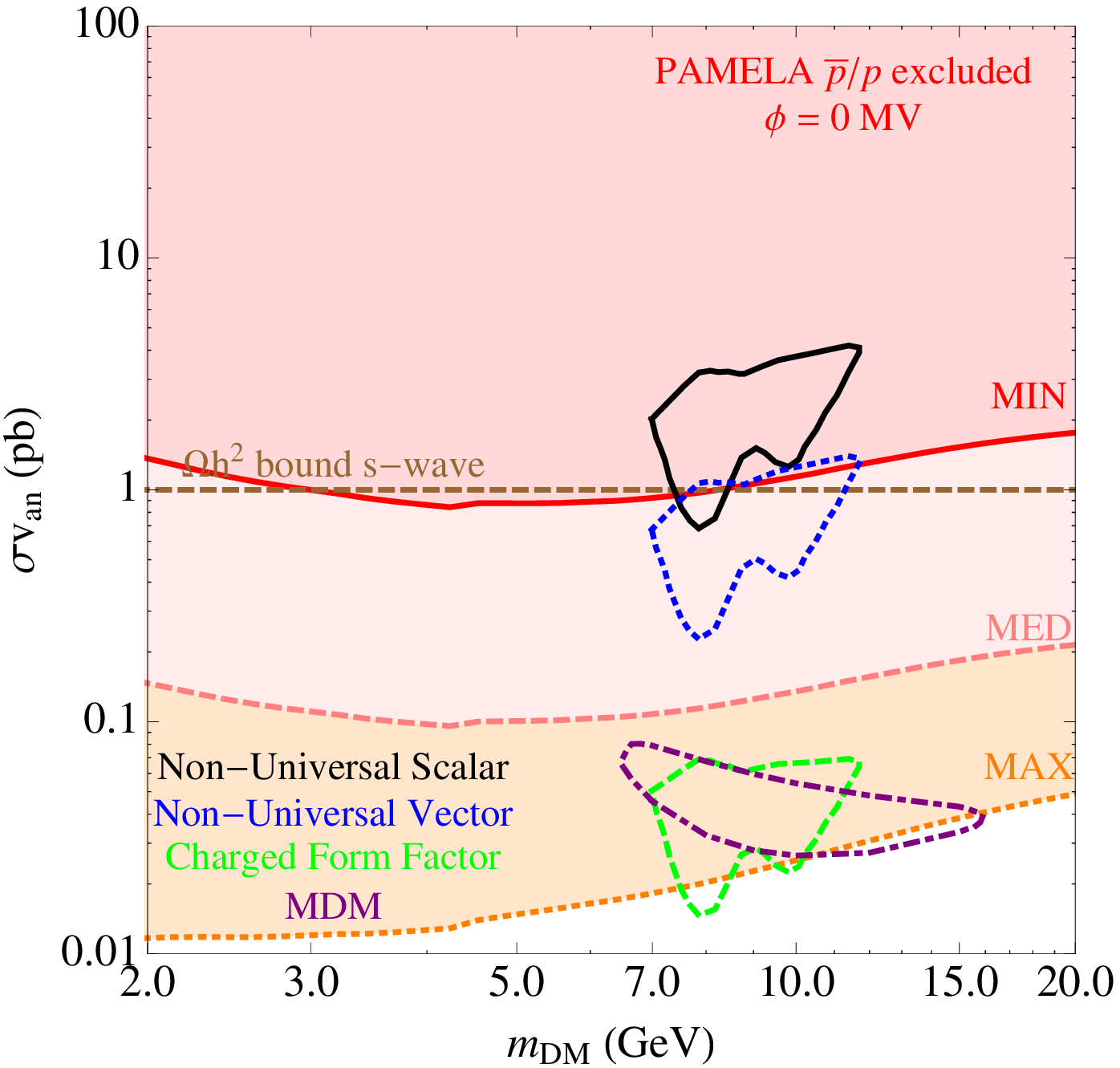} 
\includegraphics[scale=0.49]{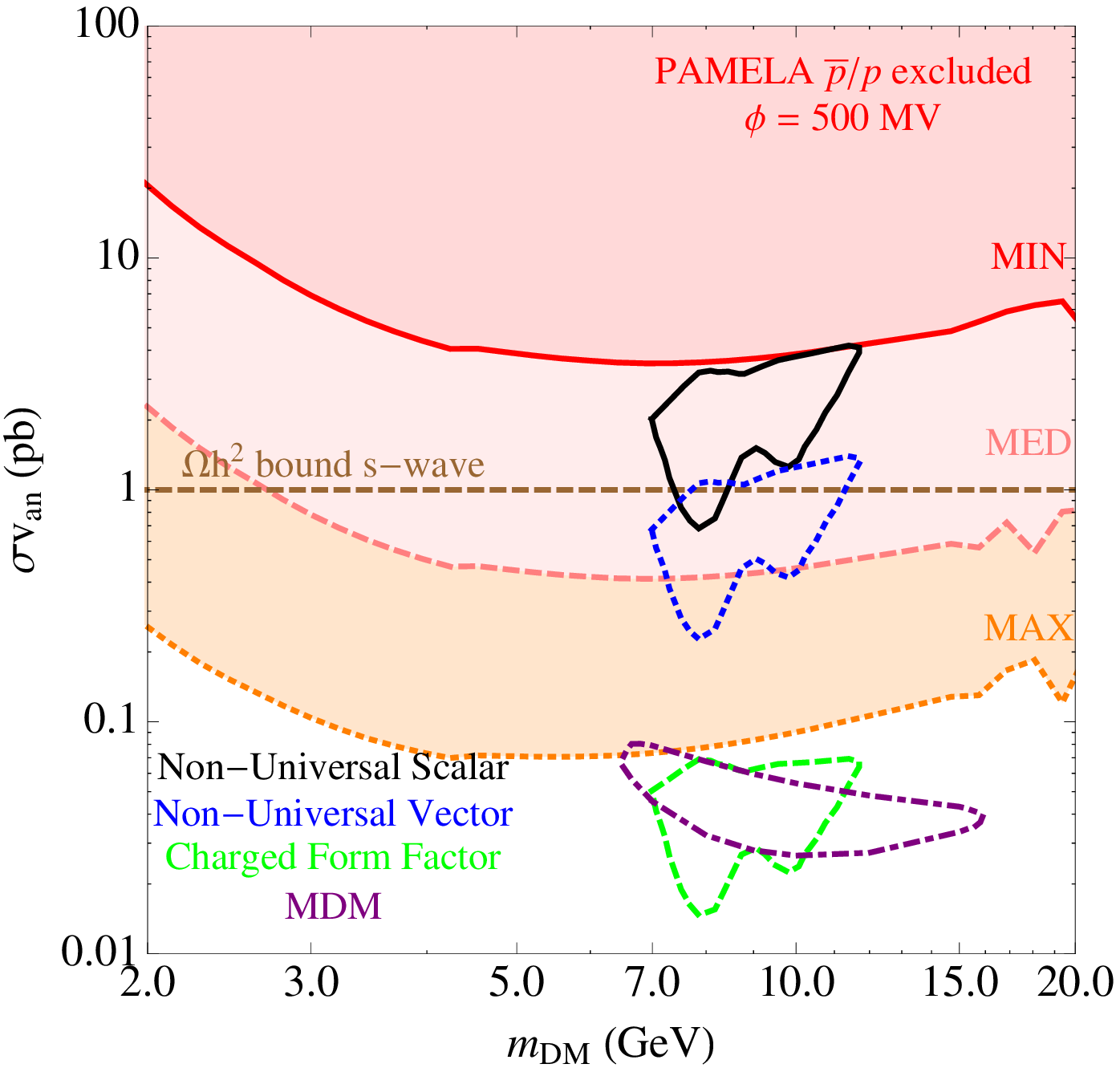} 
\caption{\em Regions disfavored by the PAMELA $\bar p / p$ spectra measurement at the 90\% C.L. for the MIN (red), MED (pink) and MAX (orange) propagation models with solar modulation potentials $\phi = 0$ MV and $\phi = 500$ MV.  Regions above the MIN limit have the most tension as no propagation model allows the $\bar p$ spectra to be consistent with the corresponding hadronic annihilation cross section.  Regions consistent with CoGeNT within the context of the non-universal scalar (solid black) and vector (dotted green) models have the most tension while the charged form factor (dashed green) and MDM (dot-dashed purple) models have little tension.  Models not listed here are well below the $\bar p$ constraints and are not shown.  Further, the limit on the hadronic annihilation cross section from relic density constraints are given for s-wave annihilators~\cite{Birkedal:2005ep}.
\label{fig:fig2} }
\end{figure} 

Given a particular propagation model, the total hadronic annihilation
rate can be constrained.  In Fig.~\ref{fig:fig2}, we present this as
colored regions which are constrained with the PAMELA data at the 90\%
C.L. for solar modulation potentials of  $\phi = 0$ MV and $\phi = 500$ MV.   In addition, we show the four largest hadronic annihilation
cross-sections, $(\sigma v)_{\rm an}$,   that are predicted from operators discussed in Section \ref{sect:II} that could explain the putative CoGeNT signal.
The contoured regions of the three models, namely the non-universal
scalar (solid black) model, the non-universal vector (dotted blue) model, and the
 charge form factor operator (dashed green) are
obtained at 90\% C.L. by translating the constraint given by
CoGeNT in Ref.~\cite{Aalseth:2010vx}. On the other hand, the contoured region of the MDM (dot-dashed purple) is
obtained as outlined in Ref.~\cite{Barger:2010gv} by fitting the recoil energy
distribution of the excess events in CoGeNT, allowing $\chi^2$
p.d.f.$<1$.
Without solar modulation, we see that effective interactions of the non-universal scalar and vector
have strong tension with the observed $\bar p/p$ flux, with the MIN
propagation model being the only allowed propagation model.  This tension is weakened somewhat by adding the solar modulation effect relevant at the solar minimum.  For the non-universal scalar, the CoGeNT-allowed region is compatible with the PAMELA $\bar p/p$ observation if at the solar minimum within the MIN model.    The charge form factor
and MDM models also contain slight tension with the $\bar p$ data, but only
within the context of the MAX propagation model.  
The models not shown in Fig. 2, the universal scalar/vector and Dirac contact models,  have
values of $(\sigma v)_{\rm an}$ that are well below the minimum range
of the figure and therefore are not at odds with the PAMELA $\bar p$
measurement.

\begin{figure}[t]
\includegraphics[scale=0.49]{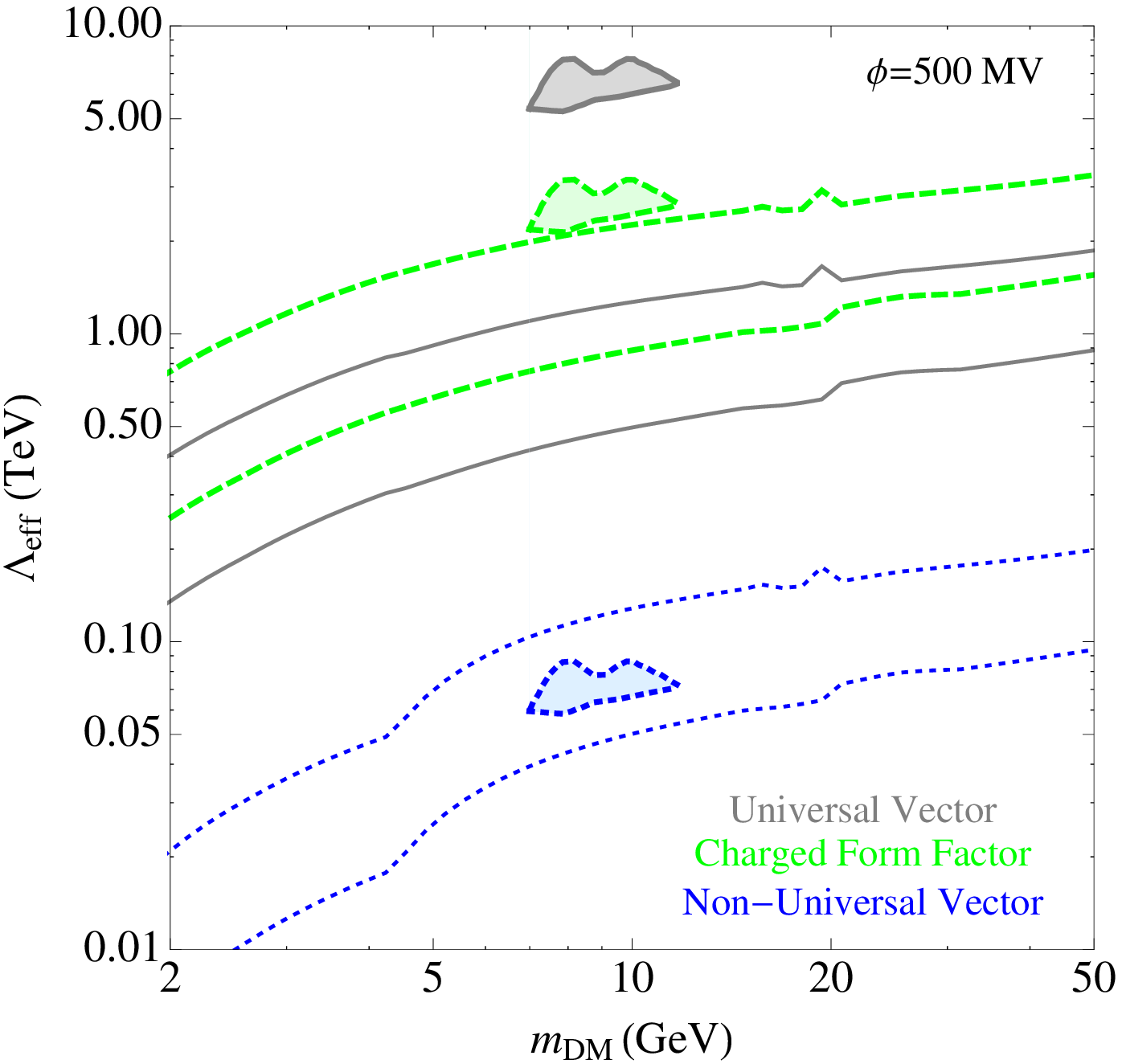}
\includegraphics[scale=0.49]{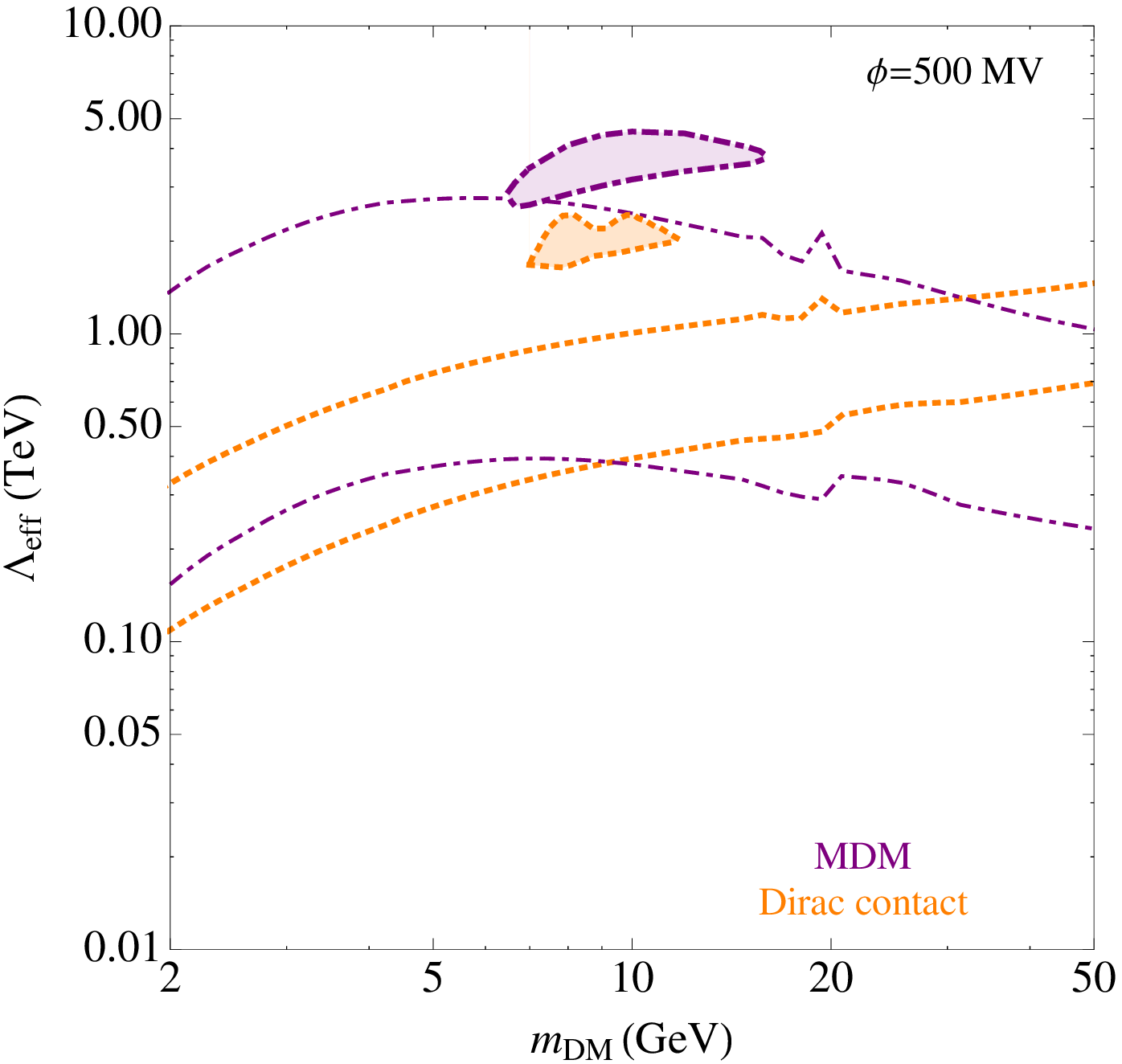}
\caption{\em The limit on the effective scale, $\Lambda_{\rm eff}$, for the charged form factor (dashed green), universal scalar/vector (solid gray) and non-universal vector (dotted blue) in the left panel and the MDM model (dot-dashed purple) and Dirac contact (dotted orange) in the right panel.  Constraints from the PAMELA $\bar p / p$ are shown by the respective curves with a $\phi=500$ MV solar modulation potential.  The PAMELA constraint for the scalar models are a factor of 3 stronger than in the vector models.  The upper curve for each model represents the limit in the MIN propagation model while the lower curve is for the MAX model.  The region above either curve is considered allowed.  The CoGeNT consistent regions for the models we consider are given by their respective filled regions. \label{fig:fig3} }
\end{figure} 

One can easily invert the dark matter annihilation cross-section to arrive at the scale of the effective interactions defined in Section \ref{sect:II}.  In Fig.~\ref{fig:fig3}, we show this scale, $\Lambda_{\rm eff}$, which is equivalent to $\Lambda_{M}$ for the MDM case and $\Lambda_{s,v,f}$ with $c_{s,v,f}=1$ and $\tilde c_{s,v}=1$ for the other models we consider.  The value of $\Lambda_{\rm eff}$ is directly related to the mass scale of the mediator particle, except for the scalar and vector cases as the dark matter mass is included in the normalization of the operators in Eqs.~(\ref{eq:os}) and (\ref{eq:ov}). In these cases the relation is dependent on the underlying short distance physics.  Most of the models are at the TeV scale or greater with the exception of the non-universal models.  This is caused by the Yukawa suppression in the scattering cross section, requiring a lighter exchange state to provide the requisite scattering rate.

Moreover, we present the range of $\Lambda_{\rm eff}$ that are constrained by the PAMELA $\bar p$ measurement.  For simplicity, we only show the MIN and MAX curves to indicate the range of constraints due to the uncertainty in the propagation model.  Values of $\Lambda_{\rm eff}$ above the band are allowed within the context of any of the propagation models we consider.  Compared to other models, the significantly broader range of the constraint in the MDM model arises from the weaker dependence on $\Lambda_{\rm eff}$,   since the MDM interaction is dimension-five, while all other interactions are dimension-six.

\section{Conclusion}
In this work we have studied the implications of the dark matter interpretations of the CoGeNT signals on the anti-proton spectrum measured by the PAMELA satellite, using the framework of effective operators. Crossing symmetry relates the scattering cross-section measured in direct detection experiments to the annihilation cross-section observed by the indirect detection experiments. Assuming a common interaction that gives rise to both the dark matter-nucleus scattering and the dark matter pair-annihilation, we have listed all relevant effective interactions of dark matter with various possible spins.  In these cases a measurement in direct detection experiments can be used to predict the outcome of indirect detections.

The absence of anomalous  $\bar p$ measurement in PAMELA, when confronted with the current CoGeNT excess of events is able to give useful constraints on the possible operators, despite the large uncertainty in the predicted anti-proton flux due to the propagation models. The most severe constraints apply to the non-universal scalar/vector interactions, where the CoGeNT parameter space is compatible with the PAMELA observation within the MIN propagation model.

Non-universal scalar interactions could arise from short distance models where the mediator particle is Higgs-like. Examples of such models which attempt to explain the CoGeNT signals can be found in Ref.~\cite{HiggsPortal}. In models where the dark matter is a complex singlet scalar, there is often mixing between CP-even scalars in the singlet and the Higgs doublet. It can be verified explicitly that such mixing does not alleviate the tension between the CoGeNT and PAMELA data. Furthermore, when one of the mass eigenstates of the CP-even scalars has a light mass not too far from that of the dark matter, the dark matter pair-annihilation through the mediator particles would be enhanced depending on the width of the mediator particles, making the PAMELA observation even more constraining. Therefore, if the MIN model does not model $\bar p$ propagation correctly, our study suggests extra ingredients may be necessary for these models to explain the CoGeNT signals without being constrained by the anti-proton measurements of PAMELA. 

\label{sect:conclusion}

\begin{acknowledgments}
We thank V.~Barger and D.~Marfatia for helpful comments and Q.-H.~Cao and H.~Zhang for earlier collaborations.  This work is supported in part by the U.S.~Department
of Energy under contracts DE-AC02-06CH11357, DE-FG02-91ER40684, and DE-FG02-84ER40173.
\end{acknowledgments}

\end{document}